\documentclass[utf8,12pt]{article}
\usepackage{sbc-template}
\usepackage[utf8]{inputenc}
\usepackage{booktabs}
\usepackage{tabularx}
\usepackage{url}
\usepackage{amsmath}
\usepackage{graphicx}
\usepackage{xcolor}
\usepackage{microtype}
\usepackage{tikz}
\usetikzlibrary{arrows.meta,positioning,calc,fit,backgrounds}
\usepackage{float}
\usepackage[cc]{titlepic}
\usepackage[hidelinks]{hyperref}
\usepackage{fancyhdr}

\newcolumntype{Y}{>{\raggedright\arraybackslash}X}
\RequirePackage{silence}
\definecolor{paFrame}{RGB}{40,54,68}
\definecolor{paSoft}{RGB}{246,248,251}
\definecolor{paAgent}{RGB}{214,228,243}
\definecolor{paBoundary}{RGB}{248,229,210}
\definecolor{paAction}{RGB}{222,236,226}

\makeatletter
\def\fnum@table{\scriptsize{TABLE~\thetable}}
\makeatother

\fancypagestyle{firstpage}{%
    \fancyhf{}%
    \fancyhead[R]{%
        \href{https://doi.org/10.5753/sbseg.2026.28577}{%
            \includegraphics[width=2cm]{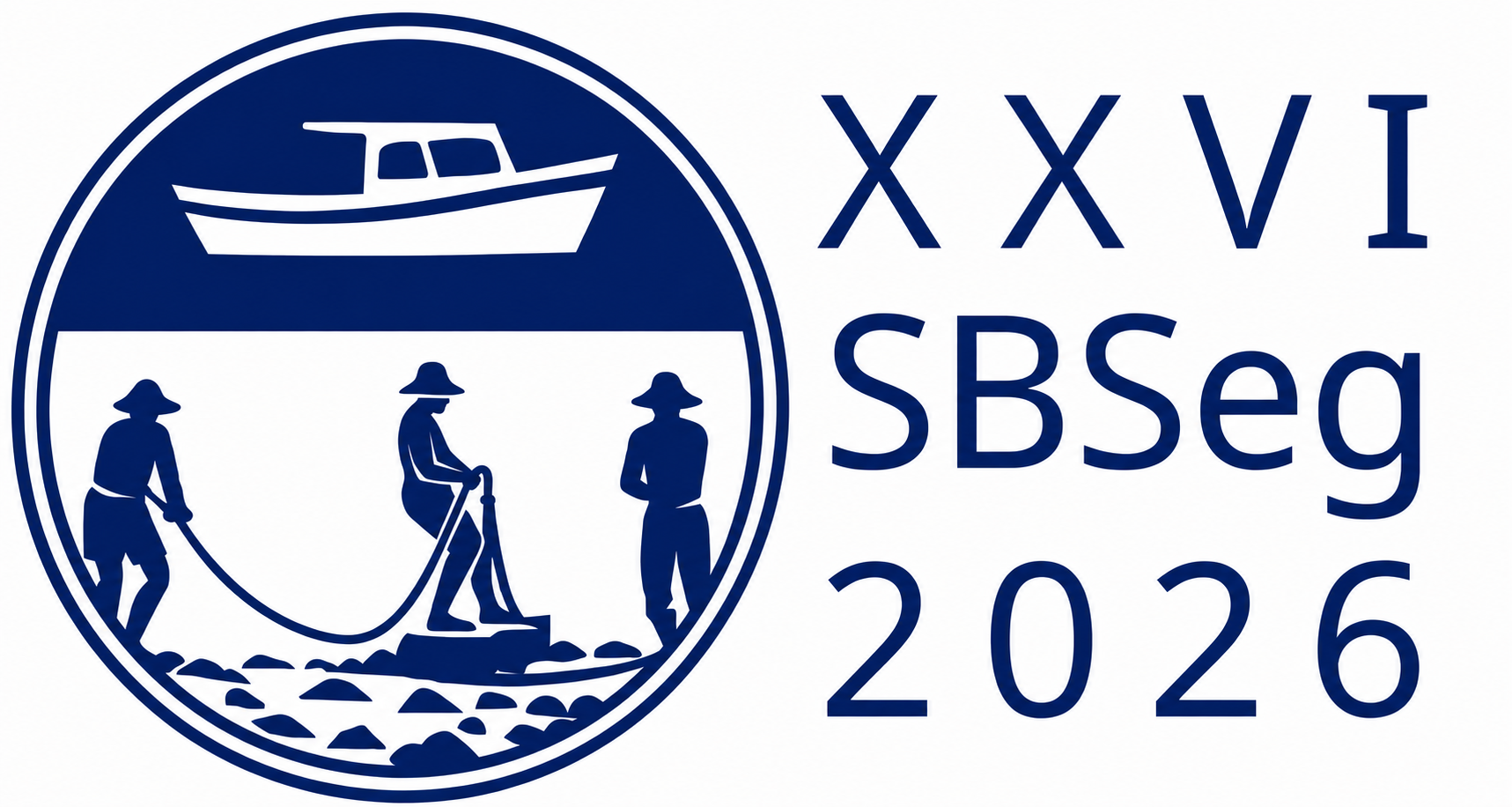}%
        }%
    }%
}

\begin{document}

\title{PocketAgents: A Manifest-Driven Library of \\Autonomous Defense Agents}

\author{Sidnei Barbieri\inst{1}, \'{A}gney Lopes Roth Ferraz\inst{1} and Louren\c{c}o Alves Pereira J\'unior\inst{1}}

\address{Divisão de Ciência da Computação -- Instituto Tecnológico de Aeronáutica
  (ITA)\\
  Praça Marechal Eduardo Gomes, 50 -- 12228-900\\
  São José dos Campos -- SP -- Brasil
\email{sidneisb@ita.br, roth@ita.br and ljr@ita.br}
}

\maketitle

\thispagestyle{firstpage}

\begin{abstract}
Connecting large language models (LLMs) to defensive enforcement requires more than asking a model whether an attack is happening. A defender must decide which model outputs may change the system state, which outputs must be rejected, and how failures should be recorded. We present PocketAgents, a manifest-driven library of autonomous defense agents. Each agent is installed as three data files: a manifest, a prompt, and a runtime context. The shared runtime gives the agent bounded telemetry access and accepts only typed reports whose requested action appears in the manifest. We implemented PocketAgents on top of Perry, a cyber-deception testbed, and evaluated two agents for the Command and Control and Exfiltration tactics in 18 closed-loop trials of a DarkSide-inspired attack on a small enterprise topology. Thirteen trials produced validated network-block actions and contained the attack; four failed schema validation; one produced a valid no-action decision. The experiments show that a typed boundary makes LLM-driven defense measurable, extensible, and attributable.
\end{abstract}


\section{Introduction}

Enterprise defense has a response bottleneck. Security operations centers (SOCs) process high volumes of alerts, rely on playbooks that analysts often adapt in practice, and still face incidents in which attackers move faster than manual triage can keep pace with. Prior SOC studies document false-positive pressure and alert ambiguity~\cite{alahmadi2022falsepositives}, \cite{yang2024socalerts}. Other studies document mismatches between tools, analyst workflows, and response playbooks~\cite{kokulu2019matched}, \cite{doplaybooks2024}. The Equifax breach highlights the operational costs of a delayed response~\cite{equifax_report}. The DarkSide campaign against Colonial Pipeline shows the same risk in ransomware response~\cite{cisa2021darkside}.

Large language models (LLMs) create a tempting shortcut: let a model inspect telemetry and recommend an action. That shortcut is unsafe if the model can directly invoke enforcement tools or if the system parses free-form prose into network changes. A recommendation such as ``block the suspicious host'' must be made concrete in terms of action, target, and scope before it changes the environment. If the output is malformed, unsupported, or ungrounded in the observed telemetry, the system should reject it and record the reason.

PocketAgents studies one architectural question: can a defensive runtime admit plug-in agents while keeping enforcement behind a deterministic boundary? The question has two operational parts: which model outputs may change the environment, and how should the runtime record the cases where an autonomous investigation fails? Perry provides the cyber-deception testbed for our closed-loop experiments~\cite{singer2025perry}; PocketAgents is the defender-side agent-library layer. Agents are installed through data files, and their outputs cross a typed enforcement boundary before any action is taken. We use specific terminology throughout: agents investigate and emit structured reports; manifests constrain what those reports may request.

The resulting abstraction is a library of defensive agents, indexed by tactical purpose and constrained by a common runtime. The current artifact exercises that idea using Command and Control (C2) and Exfiltration agents, which correspond to two tactics in the MITRE ATT\&CK (Adversarial Tactics, Techniques, and Common Knowledge) framework~\cite{mitre_attack}. They share the same runtime and differ only in their manifests, prompts, and contexts. A larger library can contain agents for other ATT\&CK tactics, such as persistence, credential access, discovery, lateral movement, or collection; this paper evaluates the interface that enables comparable evaluation of such agents. The 18-trial experiment shows that the interface is executable, that two agents can share it, and that failures become named outcome classes.

The paper makes three contributions: (1) a data-only agent package for reviewable extension when an action shape is already supported; (2) a typed admission boundary with a flat operational outcome classification for accountable enforcement; and (3) a curated closed-loop evaluation package that recomputes the reported results from preserved traces. The immediate users are researchers and defense-system builders who need to assess when an LLM-backed investigation may safely transition to enforcement.

PocketAgents changes the evaluation object. A prompt-based experiment asks whether a model can describe an attack. A rule system asks whether a predicate matches. PocketAgents asks whether an autonomous investigation can be packaged as a reusable unit, receive bounded evidence, produce an admissible report, and cause a scoped defensive action whose outcome is visible in the trace. That shift matters because the hard part of LLM-mediated defense is recognizing suspicious behavior and deciding which recognized behavior may safely be escalated to enforcement.

\section{System Model and Architecture}

PocketAgents assumes a defender-controlled execution environment. The telemetry source, dispatcher, validation code, and enforcement adapter are trusted components of the defender. The LLM backend is treated as an untrusted reasoning component whose output may be useful but cannot be executed directly. The attacker can generate network activity, trigger alerts, and influence the evidence that appears in telemetry through normal attack behavior. The attacker is not assumed to compromise the dispatcher, modify manifests, alter the validation code, or issue direct commands to the enforcement adapter.

This threat model focuses the paper on a single architectural question: can a defensive runtime admit plug-in agents while keeping enforcement behind a deterministic boundary? The present experiment measures contract compliance, containment outcome, and failure attribution under controlled attacks. The measured scope excludes prompt injection, memory poisoning, malicious tool descriptions, compromised logs, adversarial manipulation of the LLM provider, and operational false-positive rate.

The boundary shapes what these out-of-scope threats can achieve. The agent reasons from telemetry that the attacker can influence, so prompt injection via a crafted log line or a misleading process name is possible and could steer the agent toward an incorrect conclusion. The action surface remains behind the manifest: a misled agent must still emit a report that parses against the manifest schema, requests an allowlisted action, and, once grounding is enforced, names an entity present in the session telemetry. The blast radius of a successful injection is therefore bounded by the manifest. Telemetry poisoning has the same shape, since it can change which entities look suspicious while the runtime still dispatches only actions with the manifest names. Evidence sanitization and signed agent packages are the next hardening priorities.

The scope also clarifies the term ``agent.'' In this paper, an agent is a constrained investigation: a manifest defines what the agent may observe and request, a prompt defines how it investigates, and a context file defines deployment facts. The agent reasons over telemetry and emits a report; the report becomes an action only after the shared runtime validates it. This distinction is central because a library of agents should make extension easier without compromising accountability for enforcement.

The context file is part of that accountability. Enterprise traffic is rarely self-explanatory at the packet or flow level. A connection pattern resembling C2 may be an update service, a monitoring collector, an endpoint-management channel, a telemetry path, or an antivirus coordination server. A bulk transfer may be exfiltration on one host and a scheduled backup on another. PocketAgents treats context as an explicit input separated from the prompt. The runtime can carry deployment facts such as trusted subnets, host roles, user or session metadata, allowed destinations, and expected services. This is the bridge between an agent library and operational defense: the same tactical agent can run across different environments, while the manifest and context specify what counts as admissible evidence and actions.

Fig.~\ref{fig:architecture} shows how PocketAgents separates investigation, validation, and enforcement. A sensor raises a trigger, the dispatcher selects a compatible agent from the registry, binds the package to a bounded telemetry slice, and asks the agent to investigate. The investigation step is probabilistic and LLM-backed; the validation step is deterministic and performed by the dispatcher; enforcement is scoped through an adapter. The agent returns a typed report, and the typed boundary admits that report only if the manifest allows it. The dispatcher then invokes the adapter and records the outcome.

\begin{figure}[!t]
\centering
\resizebox{\linewidth}{!}{%
\begin{tikzpicture}[
  font=\footnotesize,
  box/.style={draw=paFrame, rounded corners=1.5pt, align=center, minimum height=8.4mm, inner sep=3pt, fill=white, line width=0.45pt},
  pkg/.style={box, text width=21.5mm, fill=paAgent},
  input/.style={box, text width=22mm, fill=paAction},
  run/.style={box, text width=22mm, fill=paSoft},
  gate/.style={box, text width=25mm, fill=paBoundary, line width=1.1pt},
  reject/.style={box, text width=25mm, fill=white},
  flow/.style={-Latex, line width=0.7pt, draw=paFrame},
  group/.style={draw=paFrame!70, rounded corners=2pt, inner xsep=5pt, inner ysep=8pt, fill=paSoft},
  title/.style={font=\bfseries\footnotesize, align=center}
]
  \node[pkg] (manifest) at (0,1.02) {Manifest\\{\scriptsize schema, actions}};
  \node[pkg] (prompt) at (0,0) {Prompt\\{\scriptsize procedure}};
  \node[pkg] (context) at (0,-1.02) {Context\\{\scriptsize deployment facts}};

  \node[run] (dispatcher) at (3.75,0) {Dispatcher\\{\scriptsize load package\\bind evidence}};
  \node[input] (trigger) at (3.75,1.42) {Trigger\\{\scriptsize alert class}};
  \node[input] (telemetry) at (3.75,-1.42) {Telemetry slice\\{\scriptsize bounded evidence}};
  \node[run] (agent) at (6.65,0) {LLM\\reasoning\\{\scriptsize untrusted}};
  \node[gate] (boundary) at (9.65,0) {\textbf{Typed boundary}\\{\scriptsize parse $\cdot$ allowlist\\ground target}};
  \node[run] (adapter) at (12.55,0) {Action adapter\\{\scriptsize scoped change}};
  \node[run, fill=paAction] (trace) at (15.45,0) {Environment + trace\\{\scriptsize final state\\outcome class}};
  \node[reject] (reject) at (9.65,-1.62) {Reject + class\\{\scriptsize schema, action,\\grounding, budget}};

  \begin{scope}[on background layer]
    \node[group, fit=(manifest)(prompt)(context)] (libbox) {};
    \node[group, fit=(trigger)(telemetry)(dispatcher)(agent)(boundary)(adapter)(trace)(reject)] (runtimebox) {};
  \end{scope}

  \node[title, above=1mm of libbox] {Agent library};
  \node[title, above=1mm of runtimebox] {Shared runtime};

  \draw[flow] (libbox.east) -- (dispatcher.west);
  \draw[flow] (trigger.south) -- (dispatcher.north);
  \draw[flow] (telemetry.north) -- (dispatcher.south);
  \draw[flow] (dispatcher) -- (agent);
  \draw[flow] (agent) -- (boundary);
  \draw[flow] (boundary) -- (adapter);
  \draw[flow] (adapter) -- (trace);
  \draw[flow] (boundary.south) -- (reject.north);
\end{tikzpicture}
}
\caption{PocketAgents runtime and typed enforcement boundary.}
\label{fig:architecture}
\end{figure}

An agent manifest declares the trigger class, action catalog, output schema, required fields, confirmation field, model backend, query budget, and result cap. The prompt gives the investigation procedure. The context file carries environment facts such as trusted subnets, host roles, and allowed telemetry. These files are static inputs to the runtime; they do not contain the evidence for a specific trial. The telemetry slice itself is supplied at execution time through a bounded-evidence interface.

The boundary checks a report against the manifest and the telemetry slice exposed during the session. A report is admitted only if three checks pass. First, the report must parse and contain the typed fields the manifest requires. Second, the requested action must appear in the manifest's action catalog. Third, every target entity in the report must be present in the telemetry returned during the session. The current implementation enforces the first two checks at runtime and evaluates the third after the run from recorded query results. Promoting grounding to a runtime guard is direct in principle: the runtime already records which entities the session's own queries returned, so dispatch can compare each report target against that set. Entity resolution is the substantive work because cross-host targets, ephemeral ports, and identity-level subjects require canonical forms before membership is well-defined.

The boundary uses a flat operational outcome classification, summarized in Table~\ref{tab:classes}. The labels indicate whether an operator should review evidence, fix a prompt, change a model backend, or inspect enforcement. The status column distinguishes between experimental evidence and runtime support. ``Observed'' classes appear in the 18-trial artifact. ``Runtime'' classes are implemented, but did not occur in those trials. ``Post-hoc'' marks the grounding check: the artifact measures it from recorded query results, but the runtime does not yet enforce it before dispatch.

\begin{table}[!htbp]
\centering
\caption{Boundary outcome classification.}
\label{tab:classes}
\small
\setlength{\tabcolsep}{4pt}
\begin{tabularx}{\columnwidth}{lYc}
\toprule
Class & Meaning & Status \\
\midrule
\texttt{valid\_block} & Report accepted; network block applied. & observed \\
\texttt{no\_action} & Report accepted; confirmation is false. & observed \\
\texttt{schema\_fail} & Delimiter, parse, type, or required-field check failed. & observed \\
\texttt{unsupported\_action} & Action absent from manifest catalog. & runtime \\
\texttt{ungrounded} & Target absent from session telemetry. & post-hoc \\
\texttt{budget\_exhaust} & Query or wall-clock budget exhausted. & runtime \\
\bottomrule
\end{tabularx}
\end{table}

PocketAgents is built around four design choices. The first is a data-only installation. A new agent installs without edits to the dispatcher, telemetry adapter, or enforcement adapter when its action shape is already supported. This is an assurance property: the reviewer can compare two agent directories and see whether a new behavior enters through manifest, prompt, and context files or through framework changes. In the current artifact, the C2 and Exfiltration agents satisfy this property for the same network-block action.

The second choice is typed reporting at the enforcement boundary. Tool-calling frameworks can restrict which tools a model may call; a defensive action still needs a security meaning. Blocking an Internet Protocol (IP) endpoint, isolating a host, or restoring a service must carry a target, a reason, and a scope. PocketAgents asks the model to emit a report whose fields are named in the manifest, and the dispatcher interprets only those report fields. Four Exfiltration failures are therefore rejected reports, even though the model produced natural-language analysis. Analyst-facing prose can help a human reviewer; enforcement requires an admissible report.

The third choice is bounded telemetry access. The agent receives a small interface for the evidence needed by the manifest, while the runtime withholds the rest of the environment state and shell access. This makes the execution trace inspectable. A reviewer can ask which telemetry the agent saw before a block and whether the target appears in those results. The present prototype records this evidence and uses it to compute entity precision after the run.

The fourth choice is outcome attribution. A SOC operator needs more than a containment flag. They need to know whether the system rejected malformed output, refused an unsupported action, exhausted a budget, or produced a valid no-action decision. These are different engineering problems. A schema failure can be fixed by prompt and parser work. An unsupported action asks whether the action belongs in the manifest. A no-action decision asks whether the evidence or model reasoning was insufficient. This is where PocketAgents differs from a generic ``LLM for SOC'' framing: the architecture remains useful even when the model fails, because failures are named and localized.

These choices give the admission boundary a concrete shape. The model must wrap its decision in report tags and emit the typed fields named by the manifest, including a boolean confirmation field. Structural validation checks three conditions in order: the delimiter is present, the enclosed text parses, and the required fields carry the declared types. The confirmation field separates a block from a refusal. A well-formed report whose confirmation field is false is admissible but requests no action, which is the \texttt{no\_action} outcome; a response that omits the delimiter or a required field never reaches that test and becomes \texttt{schema\_fail}. The action catalog is an allowlist, so a report that requests an action outside the manifest is rejected before dispatch. The evidence interface is bounded in the same spirit: in the prototype, the agent reads telemetry through a fixed query surface over an Elasticsearch index, and the manifest's query budget and result cap limit how much it can retrieve before it must decide. The agent therefore reasons over a delimited slice of state, and the boundary, not the model, decides whether its conclusion becomes an action.

The parser is deliberately strict, and we state its current shape plainly. It locates the report delimiter, parses the enclosed object, and verifies that the required fields are present and match their declared types. The remaining parser-hardening layer has three parts: typed enumerations and ranges, length and character limits for every field, and a delimiter that resists the injection of report-like text within the evidence. Those checks would reject malformed, oversized, or semantically odd output before any action-catalog decision is made.

\section{Instantiation}

We instantiated PocketAgents inside Perry's defender subsystem. Perry supplies emulated hosts, attack execution, telemetry capture, and result traces. PocketAgents adds the agent-library layer, including package loading, bounded LLM investigation, typed report parsing, deterministic validation, and scoped action dispatch. The scientific object is this layer and its contract; Perry remains the experiment engine.

The instantiation has a narrow boundary with the testbed. Perry remains responsible for running the scenario, collecting telemetry, and recording outcomes. PocketAgents begins when a defender-side trigger asks whether a bounded investigation should run. The runtime loads an agent package, exposes a bounded evidence interface, calls the configured LLM backend, parses the final report, and dispatches an action only after validation. This separation supports attribution. A failure before the model is called an experiment orchestration failure. A malformed report belongs to agent/model contract compliance. A validated action that does not contain the simulated attacker belongs to action or scenario semantics. The paper's taxonomy follows this boundary.

The dispatcher follows the same short path for every agent. It selects the package whose trigger matches the alert, loads the manifest, prompt, and context, binds a bounded evidence interface, and calls the configured LLM backend. The terminal response then passes through the boundary checks above. A missing report becomes \texttt{schema\_fail}; an action outside the manifest becomes \texttt{unsupported\_action}; a false confirmation becomes \texttt{no\_action}; an exhausted query or wall-clock budget becomes \texttt{budget\_exhaust}; and an accepted report dispatches the adapter and records \texttt{valid\_block}. Runtime grounding would add one guard between the action-catalog check and dispatch.

The evidence interface is the main portability hook. An agent prompt can ask for facts, but the runtime controls which queries are available and how many results are returned. In Perry, the interface is backed by experiment telemetry. In a different deployment, it could be backed by a Security Information and Event Management (SIEM) system, an Endpoint Detection and Response (EDR) system, a network-flow database, or a deception platform. The manifest names the action contract and report fields, and the runtime maps those contracts to the local environment. Perry is the evaluation substrate; PocketAgents is the reusable interface between constrained-agent investigation and defensive enforcement.

The query model in the prototype makes this interface concrete. The agent reads telemetry from a single read-only index using the Elasticsearch JavaScript Object Notation (JSON) query language and cannot write, execute, or access anything beyond that surface. The manifest sets two limits on access: a query budget caps the number of queries an investigation may issue, and a result cap caps the number of records each query returns. These limits interact with evidence sufficiency in ways the traces expose. When the relevant population exceeds the result cap, the agent switches from enumerating records to aggregating them. In the C2 trace, it groups flows by process, destination address, and port to surface the beaconing endpoint without reading every connection. The caps, therefore, do more than bound cost. They push the investigation toward summaries that fit the model's context, and an agent that cannot conclude within budget is recorded as \texttt{budget\_exhaust}; the runtime never grants unbounded reads. Each session's queries and provider exchanges are preserved in the trace, so an operator can recover the per-agent query count and wall-clock cost when deciding whether to promote an agent.

The operational separation is direct: agents investigate, while constraints admit or reject their outputs. The C2 and Exfiltration modules inspect telemetry and decide whether to emit a structured report. The manifest and dispatcher constrain those reports before enforcement. PocketAgents defines the runtime boundary that allows autonomous agents to be packaged, constrained, and evaluated as library members.

The two evaluated agents share the same runtime. The C2 agent asks whether hosts converge on a command-and-control endpoint and reports a confirmed endpoint target. The Exfiltration agent asks whether an anomalous outbound transfer is in progress and reports a destination, transfer volume, and protocol. Both agents request the same enforcement action. Moving from C2 to Exfil changes only the agent package. This is the operational definition of library membership used in the paper: a new member of the same action shape can be installed without modifying the runtime. Agents for lateral movement, credential access, discovery, persistence, or collection require different evidence and target types, but the library contract stays the same: package the investigation, bind it to local context, require a typed report, and admit only actions named by the manifest.

Table~\ref{tab:agents} makes the library contract concrete without exposing prototype internals. Both agents use the same runtime path and request the same enforcement class, but they differ in the question they ask, the confirmation gate they must satisfy, and the evidence they must produce. The Exfiltration agent records more evidence than is strictly required for admission to action. This distinction matters because an agent can produce richer evidence for analyst review without expanding the minimal enforcement predicate.

\begin{table}[H]
\centering
\caption{Evaluated agent contracts.}
\label{tab:agents}
\small
\setlength{\tabcolsep}{4pt}
\begin{tabularx}{\columnwidth}{lYY}
\toprule
Contract element & C2 & Exfiltration \\
\midrule
Question & C2 endpoint convergence & Outbound transfer anomaly \\
Confirmation & Boolean attack confirmation & Boolean transfer confirmation \\
Target & Remote network endpoint & Destination endpoint and port \\
Evidence & Host, process, cadence & Volume, direction, protocol \\
Action & Network block & Network block \\
Controls & Query budget; result cap & Query budget; result cap \\
\bottomrule
\end{tabularx}
\end{table}

The table abstracts the manifest and report schema by security role, so the reader can see each field's purpose without needing to learn prototype variable names. The authoring surface is still concrete: installing an agent means declaring the trigger class, the action catalog, the confirmation gate, the target entity, the supporting evidence fields, and the query limits as data, while the dispatcher and the enforcement adapter remain unchanged.

The target architecture also admits a coordinator above the dispatcher. Its purpose is to reconcile recommendations across scopes such as global, subnet, user, and session. Enterprise response is rarely flat: a subnet-level block, a user-session quarantine, and a global deny-list entry have different blast radii and review requirements. The current experiments involve one compatible agent per scenario and host- or network-level enforcement. This provides a concrete basis for the coordinator without making it part of the evaluation mechanism.

This scope hierarchy keeps the agent library disciplined. PocketAgents evaluates the common pattern on C2 and Exfiltration. The result is small enough to audit yet general enough to explain how additional tactical agents would enter the library without turning each one into a custom defender.

\section{Evaluation}

The evaluation uses Perry's \texttt{Darkside-EquifaxSmall} scenario. The attacker emulates scanning, lateral movement, C2 establishment at a fixed, attacker-controlled endpoint, and data exfiltration within a small enterprise topology. The defender runs one PocketAgents agent per trial. The C2 agent is triggered by evidence of command-and-control behavior; the Exfiltration agent is triggered by evidence of outbound data movement. In both cases, the only enforcement action evaluated is Perry's network-block action. The Exfiltration context declares Wazuh telemetry (ports 1514, 1515, and 55000) and OpenSearch telemetry (port 9200) as allowed services; the prompt excludes the latter ports from its initial query. The C2 trace likewise includes monitoring traffic on port 9200 alongside the malicious beacon on port 8888. These are contextual discriminators within attack trials, not benign or no-attack control trials, so they cannot estimate a false-positive rate.

The experiment was designed as a six-backend sweep: six model backends crossed with two agents and three trials per configuration, for a total of 36 attempted runs. Three backends reached the agent and defined the analysis set of 18 evaluable trials: \texttt{claude-3.5-haiku}, \texttt{gpt-4o-mini}, and \texttt{gpt-5.2}. The other three never reached the agent because the provider rejected the request before any investigation began. Two Anthropic 'latest' aliases no longer resolved and returned an application programming interface (API) \texttt{NotFoundError}; the affected families were Claude 3.5 Sonnet and Claude 3.7 Thinking. The third rejected backend was \texttt{gpt-5-mini}, which returned \texttt{BadRequestError} because the harness sent a non-default sampling temperature that this model rejects. These rejected runs pose a hazard to reproduction for anyone repeating the sweep, but recording an API rejection as an agent or model outcome would constitute a category error. The curation criterion is therefore explicit: a trial is evaluable only when the provider returns a response the agent can parse, and the run reaches Perry's terminal state. This line separates agent-and-model behavior from experiment-orchestration failure, which is why the paper reports three evaluated backends, whereas the anonymous artifact packages only the 18 evaluable trials.

The experimental unit is a complete Perry run. A trial begins when the scenario starts, continues through attack execution, alert processing, LLM-backed investigation, boundary validation, and possible enforcement, and ends when Perry records the final state. This matters because offline prompt evaluation would miss the contribution that concerns closed-loop enforcement. We inspect the final impact record alongside the runtime trace, which records the agent lifecycle and boundary outcome.

Table~\ref{tab:results} is the central result. Thirteen of 18 trials produced validated network-block actions and contained the attack at the entry host, with zero exfiltrated files. In those trials, the interval from the first LLM prompt to the first network block ranged from 26 to 76 seconds. The five non-containing trials all allowed the full chain to complete, with seven infected hosts and four exfiltrated files. The boundary explains those failures: four are \texttt{schema\_fail}, where the model produced an analysis but not the required structured report object; one is \texttt{no\_action}, where the report was valid but did not confirm malware. The table reports the three outcome classes in separate columns, the mean block time across block trials, and the number of trials per block. The upper panel shows per-agent and per-backend counts, while the lower panel aggregates by backend. This system diagnoses whether a backend fails at contract compliance or at the decision itself.

\begin{table}[H]
\centering
\caption{Closed-loop trial outcomes.}
\label{tab:results}
\footnotesize
\setlength{\tabcolsep}{3pt}
\begin{tabularx}{\columnwidth}{Yrrrrc}
\toprule
Configuration & Block & Schema & No action & Time (s) & Contained \\
\midrule
C2 + \texttt{claude-3.5-haiku}    & 3 & 0 & 0 & 46.0 & 3/3 \\
C2 + \texttt{gpt-4o-mini}         & 2 & 0 & 1 & 50.0 & 2/3 \\
C2 + \texttt{gpt-5.2}             & 3 & 0 & 0 & 32.0 & 3/3 \\
Exfiltration + \texttt{claude-3.5-haiku} & 2 & 1 & 0 & 75.0 & 2/3 \\
Exfiltration + \texttt{gpt-4o-mini}      & 0 & 3 & 0 & --   & 0/3 \\
Exfiltration + \texttt{gpt-5.2}          & 3 & 0 & 0 & 34.7 & 3/3 \\
\midrule
\multicolumn{6}{l}{\emph{By backend (both agents combined, six trials each)}} \\
\texttt{claude-3.5-haiku} & 5 & 1 & 0 & 57.6 & 5/6 \\
\texttt{gpt-4o-mini}      & 2 & 3 & 1 & 50.0 & 2/6 \\
\texttt{gpt-5.2}          & 6 & 0 & 0 & 33.3 & 6/6 \\
\midrule
\textbf{All 18 trials} & \textbf{13} & \textbf{4} & \textbf{1} & \textbf{45.2} & \textbf{13/18} \\
\bottomrule
\end{tabularx}
\end{table}

The trace files support the interpretation. In a successful C2 trial, the runtime records agent admission, the first prompt, and the network block against the C2 endpoint; the final impact record shows one infected host and no exfiltration. In a failed Exfiltration trial with \texttt{gpt-4o-mini}, the runtime records that the model did not emit the required typed report for either of the investigated web servers; the final impact record shows 7 infected hosts, 4 exfiltrated files, and no blocks. The anonymous data-and-analysis package contains the 18 canonical trials, Secure Hash Algorithm 256 (SHA-256) hashes for the final impact records and runtime traces, and scripts to recompute the outcome distribution.

For each trial, we record the boundary outcome, the number of infected hosts at termination, the exfiltrated files, the time to enforcement or terminal failure, and the target precision for any enforced block. In the 13 \texttt{valid\_block} trials, the final impact record shows one infected host and no exfiltrated files. In the five non-containing trials, there are seven infected hosts and four exfiltrated files. Target precision for applied blocks is 1.0: each enforced IP/port pair appears in the trace as the attacker-controlled C2 endpoint. This is a measurement over the recorded traces; runtime grounding would turn the same check into an admission guard.

The latency numbers are useful as experimental traces. Eight C2 blocks occur in 26 to 62 seconds after the first prompt; five Exfiltration blocks occur in 31 to 76 seconds. Perry has already raised the trigger by the time the prompt is sent, so these values measure the agent-runtime segment: bounded evidence access, model response, report parsing, validation, and action dispatch. Their role is to show that the runtime is closed-loop and executes before the experiment reaches its final impact state.

The C2 agent is easier than Exfiltration in this experiment. Eight of nine C2 trials contained the attack, and the remaining trial ended in a valid no-action report. Exfiltration contained five of nine trials and produced all four schema failures. The difference is informative because the two agents share the same runtime and action path. C2 has a compact target, the attacker-controlled endpoint, and the action target maps directly onto that endpoint. Exfiltration asks the model to connect volume, direction, host role, and protocol before emitting the same action. The harder case, therefore, stresses the report contract rather than the enforcement adapter.

Model behavior should also be read through the lens of the boundary. \texttt{gpt-5.2} produced validated blocks in all six evaluated trials. \texttt{claude-3.5-haiku} produced five validated blocks and one schema failure. \texttt{gpt-4o-mini} produced two validated blocks, one valid no-action decision, and three schema failures. The value of this comparison is diagnostic: it indicates whether a backend meets the enforcement contract, violates the report schema, or returns a valid decision declining action. A model response that produces prose instead of a report changes in the environment state; the runtime records it as a rejected enforcement attempt.

The boundary evaluates the terminal response, and the traces show why this distinction carries weight. All four schema failures reach the same terminal condition: the final response contains no parseable \texttt{\textless report\textgreater} object. The two backends arrive there by different paths. In the three \texttt{gpt-4o-mini} Exfiltration trials, the model produces natural-language analysis of each investigated web server but never emits a report object, so the dispatcher raises a contract error for both candidates. The single \texttt{claude-3.5-haiku} Exfiltration failure is more subtle: the model emits eight intermediate report objects across the session, yet its terminal turn omits the report and produces the same condition. The boundary treats the two cases identically. Intermediate reasoning, and even intermediate well-formed reports, do not authorize an action; only an admissible terminal report does. From the runtime's perspective, a model that almost complied is indistinguishable from one that never did. Accepting the last well-formed intermediate report would relax this rule, but it would also risk enforcing on evidence the agent itself superseded, so the boundary requires the terminal decision.

The single no-action outcome is a different event, and the contrast is the core of the taxonomy. In the \texttt{C2}/\texttt{gpt-4o-mini} trial classified as \texttt{no\_action}, the model emits a valid terminal report indicating that malware was not confirmed, and no block follows. The same backend thus spans all three observed classes, as the per-backend panel of Table~\ref{tab:results} shows: it satisfies the contract on the simpler C2 agent, declines once when the evidence does not confirm an attack, and violates the contract on the harder Exfiltration agent. Outcomes, therefore, vary with the interaction between the agent and the contract, and with the boundary records that indicate which case occurred. This is the operational payoff of typed admission: the same event that a binary ``did it block?'' view would record as a single failure is resolved into a missing terminal report, a grounded refusal, and a contained block, each with a different repair path.

The execution times clarify what the runtime measured. The mean runtime includes scenario execution, attack progress, alert processing, LLM interaction, and enforcement; it is a scenario-level duration that accounts for model latency, overhead, and experimental time. In the curated dataset, mean execution time ranges from roughly 224 seconds for Exfiltration with \texttt{gpt-4o-mini} to roughly 949 seconds for C2 with \texttt{gpt-5.2}. The shorter value reflects a full compromise across all three Exfiltration \texttt{gpt-4o-mini} trials, as each ended in a schema failure. For a defensive system, the useful metric is a validated action that arrives before the attacker reaches its objective. This is why the paper reports the outcome class together with infected hosts and exfiltrated files.

Two caveats define how the dataset should be read. First, the same scenario is repeated three times per configuration, so the trials measure run-to-run variation in the experiment and model interaction more than diversity in attacker behavior. Second, action success is tied to Perry's simulated network-block semantics. Within that scope, the evidence is strong for architectural accountability: the framework reaches the enforcement adapter, applies the action, records the outcome, and assigns non-containment to named boundary classes.

The artifact is organized around the evidence used in the paper. It preserves the 18 evaluable trials used in the quantitative result and documents the excluded provider-configuration failures as experiment-orchestration hazards, separate from agent outcomes. Each canonical trial contains Perry outputs, the final impact record, the LLM interaction trace, and hash metadata. A recomputation script recovers the 13/4/1 outcome distribution, per-backend containment rates, and time-to-block intervals directly from those traces. Because hosted model endpoints change, the primary reproducibility target is trace-based recomputation from the preserved runs. Full live re-execution of Perry and paid LLM calls remains a separate mode of reproduction.

\section{Discussion}
\label{sec:discussion}

The experiments show why the library boundary matters. PocketAgents make each agent a package whose authority is declared in a manifest, whose evidence is provided through a bounded interface, and whose output becomes an action only after validation. In the artifact, this makes C2 and Exfiltration comparable library members: both share the runtime and action path but ask different questions of the telemetry.

The main systems insight is that defensive action is context-dependent. A flow can resemble C2 traffic because of destination, cadence, or service role, but the same raw shape can also appear in management telemetry, update infrastructure, monitoring agents, or antivirus services. Exfiltration has a similar ambiguity: high-volume outbound transfer is suspicious only relative to host role, protocol, destination, and expected behavior. PocketAgents makes that context explicit by separating the static agent contract from the runtime context and bounded evidence. The action boundary then checks the model's conclusion against the report shape and the action scope admitted for that agent.

One trace makes the context argument concrete. In the C2/\texttt{gpt-5.2} run, a Falco rule flags a web server as suspicious, the dispatcher admits the C2 agent, and the agent issues Elasticsearch queries over the host's telemetry. The aggregated results show two flows to the same external host, 192.168.1.225: the monitoring stack on port 9200, the benign log index, and a process named \texttt{splunkd} beaconing on port 8888. The agent confirms the second as command-and-control, emits a terminal report naming that endpoint, and the boundary admits a single network block. The impact record shows the attack contained at the entry host, with no exfiltrated files and no captured flags. This is the separation the architecture is built for: a single destination address carries both benign and malicious traffic, so the decision turns on the process and port, not the address alone. A static rule keyed to that address would either block the monitoring path or miss the beacon; the agent reports the endpoint implicated by the telemetry, and the boundary records the block against it.

The outcome classification turns model behavior into an operational state. A valid block is a traceable enforcement event. A schema failure is a rejected command with a repair path. A no-action report is a valid decision that can be inspected against the evidence. This is the value of the 13/4/1 result: the runtime separates non-containment into named cases and tells the defender where the chain broke.

The current artifact demonstrates six mechanisms together: package-level extension, typed admission, closed-loop containment, context separation, model sensitivity, and trace-based reproducibility. The central result is the combination: the same runtime records an accepted block, a valid no-action decision, and schema rejection as different system states with different repair paths.

The evaluated agents also show why a library can be organized by tactic while keeping tactics outside the enforcement mechanism. MITRE ATT\&CK supplies a useful indexing vocabulary: C2 and Exfiltration correspond to different defender questions, evidence needs, and report fields. The enforcement runtime admits an action if a report satisfies a manifest; the tactic label indexes the library but carries no authority on its own.

The evaluated scope isolates the interface: two tactics, one scenario family, and one network-block action. The artifact shows that agents can be packaged as data, executed in a shared runtime, rejected deterministically when the report contract fails, and tied to impact records when enforcement succeeds. Broader traffic mixes, additional topologies, and action surfaces such as host isolation, session termination, credential revocation, or deception reconfiguration test breadth. The current result establishes the typed, traceable boundary that makes those extensions comparable.

The broader lesson is that autonomous defense needs a systems boundary before it needs a larger prompt. PocketAgents turns the usual questions about LLM-mediated defense into artifacts: the package states what the agent may do, the trace states what evidence it saw, the report states what it requested, and the outcome class states how the runtime handled the request. The experiments provide a concrete demonstration of this claim: two tactical agents, a shared runtime, 18 complete executions, and a distribution of results that distinguishes containment, malformed reports, and valid no-action behavior.

\section{Threats to Validity}

Five threats bound the claims. The first is causal attribution. The classification shows that the typed boundary partitions non-containment into named classes. The study omits a free-form baseline in which the same model outputs are parsed without the report contract, so the claim is failure attribution, not improvement in raw model decision quality. Isolating the boundary's marginal effect requires a baseline comparison.

The second is the single prompt per agent. Each agent carries one fixed investigation prompt, so the per-backend differences in Table~\ref{tab:results} are entangled with how well that prompt fits each model. A backend that fails the report contract under our prompt might satisfy it under a different one. The result measures contract compliance under fixed prompts; backend capability ranking belongs in a separate benchmark.

The third concerns the measured quantities. The block-time measurement covers the agent-runtime segment after Perry raises the trigger, not end-to-end detection, and target precision is computed after the run rather than enforced during admission. Both are trace measurements, not runtime guarantees. Runtime grounding would convert a target precision measurement into an enforced invariant.

The fourth is false-positive behavior. The experiments are attack trials and omit benign or near-benign controls. They show how the boundary handles malicious activity and malformed agent outputs; they do not estimate how often an agent would over-block during ordinary operations. This matters because an endpoint block can affect shared services, update paths, or monitoring channels. PocketAgents limits the blast radius by requiring a named action and target shape, yet it does not currently support richer approval conditions, such as role-based exceptions, time-bounded blocks, or multi-party review for shared infrastructure. A benign workload is therefore the next experiment needed to measure specificity and operator burden.

The fifth is statistical scope. With three runs per configuration within a single scenario family, the counts describe run-to-run behavior of this attack, not an attack distribution, and they do not support confidence intervals. The next study should enforce grounding at dispatch, add benign and mixed-traffic controls plus a free-form baseline, and then extend the sweep to dated model identifiers and additional scenarios; the artifact preserves the traces needed for that expansion.

\section{Related Work}

PocketAgents sits at the point where an agent's conclusion crosses into defensive enforcement. Related systems place boundaries around network control, observed provenance, tool execution, incident response, or benchmarked agent behavior. The key question is where each system assigns authority.

Programmable security control planes make enforcement explicit. SANE, Kinetic, and Precise Security Instrumentation (PSI) express network and access decisions over deterministic substrates~\cite{casado2006sane,kim2015kinetic,yu2017psi}. PocketAgents keeps the same systems discipline while changing the source of the decision: the conclusion comes from an LLM-backed investigation. The deterministic part, therefore, moves to the boundary around the model's output. A report has to parse, name an allowed action, and identify a target shape before the runtime can dispatch an adapter.

Provenance systems establish the complementary requirement that security decisions remain tied to evidence. UNICORN and NoDoze use system provenance to connect alerts to the causal activity that produced them~\cite{han2020unicorn,hassan2019nodoze}. Agent-Sentry introduces a similar approach for LLM agents with tools by building execution-provenance graphs, learning benign action envelopes, checking sensitive arguments, and invoking an intent judge for ambiguous actions~\cite{sequeira2026agentsentry}. PocketAgents uses a declared boundary: the manifest specifies which report fields and actions are admissible, and the trace records the agent's queries. The two approaches are compatible. A provenance monitor can determine whether a tool action is unusual; PocketAgents can determine whether a defensive conclusion has the security shape needed to affect the environment.

Recent LLM-agent safety systems constrain agents at lower levels of execution. AgentSpec enforces rules at the tool-call layer~\cite{agentspec2026}, and IsolateGPT isolates execution across trust domains~\cite{wu2025isolategpt}. Top-venue security work on web-enabled LLMs and proactive defenses against LLM agents shows why these boundaries matter once a model can browse, call tools, or interact with a deceptive environment~\cite{kim2025webllms,ayzenshteyn2025cloak}. These systems limit what an agent can call or access. PocketAgents assigns meaning to the decision that remains after investigation: the report must include a target, a reason, a confirmation gate, and an allowed action. This layer matters for defense because a safe tool call can still be the wrong defensive action if it targets a shared service, the wrong host, or an entity absent from the evidence.

Incident-response work treats safety as a lifecycle. AIR defines a domain-specific language for detecting incidents, guiding containment and recovery, and synthesizing future guardrail rules inside the agent loop~\cite{xiao2026air}. Empirical SOC studies show why such a structure is needed: analysts face false-positive pressure, mismatched tools, playbook drift, and alert-triage work that detection alone does not solve~\cite{alahmadi2022falsepositives,yang2024socalerts,kokulu2019matched,vermeer2023alertalchemy,doplaybooks2024}. PocketAgents supplies a narrower systems primitive for these workflows: a typed, attributable enforcement record that a SOC process can route, audit, escalate, or reject.

Benchmarks for agent safety show why aggregate success rates are insufficient. Agent-SafetyBench evaluates agents in interactive environments and failure modes, identifying broad safety gaps among the tested agents~\cite{zhang2025agentsafetybench}. PocketAgents turns that principle into an operational classification for security runtimes. A failed trial becomes \texttt{schema\_fail}, \texttt{no\_action}, \texttt{unsupported\_action}, \texttt{budget\_exhaust}, or, once runtime grounding is implemented, \texttt{ungrounded}; each label names a different repair path.

Offensive-agent results set the risk bar. PentestGPT shows that planning, memory, and tool use can produce effective attack behavior in security tasks~\cite{deng2024pentestgpt}. The defensive side has the harder contract: an attacker can benefit from partial progress, while a defender must explain which autonomous conclusion is allowed to change the system state. PocketAgents addresses that crossing point.

\section{Conclusion}

PocketAgents makes explicit the boundary that autonomous defense systems must get right: the point where an LLM-backed investigation becomes a state-changing defensive action. Its agent package fixes the investigation surface in data, its runtime admits only typed terminal reports, and its trace records whether each run produced an action, a malformed report, or a valid refusal. In 18 closed-loop Perry trials with C2 and Exfiltration agents, this boundary produced 13 validated blocks, 4 schema rejections, and 1 valid no-action decision. The result gives agentic defense a sharper evaluation object: each attempted enforcement becomes an auditable system state, reproducible from traces, comparable across agents, and extensible without granting the model direct authority over the environment.

\section*{Artifact Availability}

The public data-and-analysis artifact is available at \url{github.com/sidneibarbieri/PocketAgents}; the evaluated release is tagged \texttt{v1.0.1}. It contains the curated trial set, original Perry outputs, SHA-256 hashes for final impact records and runtime traces, reference agent packages, and scripts that regenerate the main quantitative results reported in the paper. Code and documentation are released under MIT, and the preserved experimental dataset is released under CC BY~4.0.

\section*{Acknowledgments}

This work is funded by FAPESP \#2020/09850-0, \#2024/21006-1, \#2022/00741-0; CNPq 312294/2025-5, 409743/2025-9; and CAPES.

\bibliographystyle{sbc}
\bibliography{references}

\end{document}